\documentclass[12pt]{article}
\usepackage{geometry}
\usepackage{a4}
\usepackage{graphicx}
\usepackage{epsf}
\usepackage{amsmath}
\usepackage{amssymb}
\usepackage{cite}
\newcommand{\be}{\begin{equation}}
\newcommand{\ee}{\end{equation}}

\newcommand{\Rmnum}[1]{\expandafter\@slowromancap\romannumeral #1@}
\newcommand{\bea}{\begin{eqnarray}}
\newcommand{\eea}{\end{eqnarray}}

\begin{document}
\def\A{{\mathbb{A}}}
\def\B{{\mathbb{B}}}
\def\C{{\mathbb{C}}}
\def\R{{\mathbb{R}}}
\def\s{{\mathbb{S}}}
\def\T{{\mathbb{T}}}
\def\Z{{\mathbb{Z}}}
\def\W{{\mathbb{W}}}
\begin{titlepage}
\title{Geometric Critical Exponents in Classical and Quantum Phase Transitions}
\author{}
\date{
Prashant Kumar, Tapobrata Sarkar
\thanks{\noindent E-mail:~ kprash, tapo @iitk.ac.in}
\vskip0.4cm
{\sl Department of Physics, \\
Indian Institute of Technology,\\
Kanpur 208016, \\
India}}
\maketitle
\abstract{
\noindent
We define geometric critical exponents for systems that undergo continuous second order classical and quantum phase transitions. These relate scalar quantities on the information
theoretic parameter manifolds of such systems, near criticality. We calculate these exponents by approximating the metric and thereby solving geodesic equations analytically, 
near curvature singularities of two dimensional parameter manifolds. The critical exponents are seen to be the same for both classical and quantum systems that we
consider, and we provide evidence about the possible universality of our results. 
}
\end{titlepage}
\section{Introduction}
Methods of information geometry (IG) provide valuable insights into the physics of phase transitions. The starting point here is the definition of a Riemannian metric tensor in the
space of parameters describing systems that undergo such transitions. The metric on the parameter manifold (PM) can either be induced from an equilibrium 
thermodynamic state space \cite{rupp} or from the Hilbert space structure of quantum states \cite{pv}. For classical phase transitions (CPTs), the coordinates 
characterizing the PM can be thermodynamic variables (and suitable Legendre transforms thereof) while for quantum phase transitions (QPTs), these
are naturally defined by the coupling constants of the theory that appear in its Hamiltonian. 

In a geometric context, scalar quantities in the parameter manifold of systems undergoing phase transitions are important objects to study. These provide global 
(i.e coordinate independent) characterizations of the manifold. Consider a two dimensional PM with coordinates $x^{\mu}$ and 
metric $g_{\mu\nu}$ $(\mu,\nu=1,2)$, and an invariant line element $d\lambda^2 = g_{\mu\nu}dx^{\mu}dx^{\nu}$. 
The scalar curvature (or Ricci scalar) $R$ constructed out of the metric via a standard formula \cite{curvature} fully characterizes the 
curvature properties of the manifold. Scaling relations involving the curvature scalar have been established. 
For CPTs, it can be shown via Gaussian fluctuation theory that $R \sim \xi^d$, with $\xi$ the correlation length and $d$ the system dimension 
\cite{rupp},\cite{tapo1},\cite{tapo2}.\footnote{The definition of the scalar curvature \cite{curvature} has a sign ambiguity, depending on two different ways to contract the 
Riemann tensor. Thus, more appropriately, we have the relation $|R| \sim \xi^d$. However, in order to simplify the notation, we will not write this explicitly in what follows.}
Covariant thermodynamic fluctuation theory \cite{rupp} provides excellent justification for this result, and this fact is also corroborated by explicit computations
in several model systems. However, note that 
in general there is no such simple relation for QPTs. For example, in the transverse XY spin chain, 
$R \propto (1-e^{-1/\xi})^{-1}$ in part of the anisotropic region \cite{tapo3}, i.e only near the anisotropic transition line, $R \sim \xi$.

In general, in a geometric setup, it is natural to express scaling relations in terms of geometric invariants. That is, we look for relationships of the form 
${\mathcal A} \sim {\mathcal B}^a$ where both ${\mathcal A}$ and ${\mathcal B}$ are scalar quantities on the PM and for such a relationship near the critical point, 
we will call ``$a$'' a geometric critical exponent. As an example, given the invariants $R$ and $\lambda$, we can envisage a critical scaling relation of the 
form $R \sim \lambda^a$. Although obtaining such relations might not be feasible in general, we will outline a method by which this is possible near criticality. 

The basis of our construction is to study geodesics on the parameter manifolds for classical and quantum phase transitions, near criticality. Geodesics, which are analogues
of straight lines on curved spaces, are paths that minimize distances on a possibly curved manifold. 
The physical meaning of a non-local geodesic distance can be gleaned in CPTs : generalising an argument of \cite{rupp}, a large distance between two points in the 
parameter space means a small probability that these are connected by a fluctuation. Although a corresponding statement in zero temperature QPTs is somewhat unclear, geodesics are
nonetheless extremely important objects to study in any geometric setup. 

If $x^{\mu}$ denotes the coordinates on a curved manifold, then a geodesic on the manifold satisfies 
${\ddot x^{\mu}} + \Gamma^{\mu}_{\nu\rho}{\dot x^{\nu}}{\dot x^{\rho}} = 0$
where $\Gamma^{\mu}_{\nu\rho}$ are the Christoffel connections \cite{curvature} and the overdot indicates a derivative with respect to an affine parameter along the geodesic,
which is taken to be the square root of the line element $\lambda$. We will define $u^{\mu} = {\dot x^{\mu}}$, the ``velocity'' vector tangent to the 
geodesics, and consequent to our choice of the affine parameter, these will be normalized to $u^{\mu}u_{\mu}=1$. This follows from the fact that along an affinely parametrized
geodesics, the quantity $u^{\mu}u_{\mu}$ is a constant, and can be set to unity (see section 1.3 of \cite{poisson}). From a practical point of view, it is useful to note that geodesics 
can be obtained from a variational principle, from a Lagrangian ${\mathcal L} = \frac{1}{2}\left(g_{\mu\nu}{\dot x^{\mu}}{\dot x^{\nu}}\right)$, a fact that will be important for us later. 
In general, for the IG of a given system, geodesics are given by
coupled second order non-linear differential equations, and need to be solved numerically using appropriate boundary conditions. They show interesting behavior 
in regions where the scalar curvature of the parameter manifold diverges \cite{tapo4}. We will be interested only in geodesics that reach the critical point. In special cases, analytic solutions to such 
geodesic equations are possible, and when these are inverted so as to solve for the coordinates in terms of the affine parameter, we will obtain 
geometric exponents of the form $R \sim \lambda^a$.

In fact, we can go further. As is known in the literature, curvature effects on geodesics can be obtained by considering a collection of non-intersecting geodesics 
(called a congruence), which can be treated analogous to deformable fluids. The evolution of such a geodesic congruence can be naturally specified in terms of three scalar parameters, 
namely the expansion, shear and rotation (collectively called the ESR parameters). Consider a collection of non-intersecting geodesics on a two dimensional parameter manifold. 
One can imagine a vector $\xi^{\mu}$ joining two points on nearby geodesics, called the deformation vector. Its variation with the affine parameter characterizes the evolution of the congruence. 
Introducing the tensor $B^{\mu}_{\phantom{1}\nu}=\nabla_{\nu}u^{\mu}$, it can be checked by geometric operations \cite{poisson},\cite{sayan1} that 
$\nabla_{\nu}\xi^{\mu}u^{\nu} = B^{\mu}_{\phantom{1}\nu}\xi^{\nu}$ i.e $B^{\mu}_{\phantom{1}\nu}$
is a tensor that measures the amount by which $\xi^{\mu}$ fails to get parallel transported along the congruence, and hence captures information about the curvature
of the parameter manifold. Here, $\nabla_{\mu}$ is a covariant 
derivative on the PM, defined on a generic vector $V^{\mu}$ by $\nabla_{\mu}V^{\nu} = \partial_{\mu}V^{\nu} + \Gamma^{\nu}_{\mu\lambda}V^{\lambda}$.
$B_{\mu\nu}$ is called the evolution tensor. By a standard procedure in matrix algebra, it can be decomposed into the following irreducible parts:
\begin{equation}
B_{\mu\nu}=\theta h_{\mu\nu}+\sigma_{\mu\nu}+\omega_{\mu\nu}
\label{evolution}
\end{equation}
where
$h_{\mu\nu}=g_{\mu\nu}-u_{\mu}u_{\nu}$
is the projection tensor and 
\begin{eqnarray}
&~&\theta = B^{\mu}_{\phantom{1}\mu},~~~\sigma_{\mu\nu}  = \frac{1}{2}\left(B_{\mu\nu}+B_{\nu\mu}\right)-\theta h_{\mu\nu}~,\nonumber\\
&~&\omega_{\mu\nu} = \frac{1}{2}\left(B_{\mu\nu}-B_{\nu\mu}\right).
\label{esr}
\end{eqnarray}
$\theta$, $\sigma^2$ and $\omega^2$ are the ESR parameters respectively for the geodesic congruence. In time dependent situations, they specify the 
expansion, shear and rotation of the deformable fluid as it flows along the congruence. For our case, these are scalar parameters that signify changes in
the shape and size of the geodesic congruence. For two dimensional parameter manifolds, $\sigma_{\mu\nu}$ and $\omega_{\mu\nu}$ are identically zero, and $\theta$ is the 
only parameter that characterizes the congruence. Generally, $\theta$ diverges as a congruence approaches a singularity, i.e all the geodesics converge to a point \cite{geofocus}. 
In our case, the geodesic congruences are simply collections of lines on the curved parameter manifold, and $\theta$ exhibits similar interesting features near a curvature singularity. 
$\theta$ is a scalar and is also related to the scalar curvature of the manifold, as we will elaborate upon later. For the moment, we note that if we obtain $\theta$ 
as a function of the affine parameter near a curvature singularity of the manifold,
we would obtain a second geometric exponent. The rest of this paper is devoted to understanding the behavior of $R$ and $\theta$ as functions of $\lambda$.

As alluded to before, the main problem at hand is that geodesic equations do not, in general, admit analytic solutions and have to be solved numerically with appropriate
boundary conditions. This will not be very useful for us,\footnote{We believe that numerical results should be possible to obtain in generic examples, although in
this paper, we will only consider models in which analytical solutions are possible} and 
to circumvent the problem, we adopt the following strategy. Since we are mainly interested in regions of the PM close to second order phase transitions,
we approximate the metric on the manifold at the critical point, while not losing any information about the scalar curvature. After this, we rewrite the metric in a transformed set of 
coordinates so as to remove the dependence on one of them. This greatly simplifies the geodesic equations, while introducing a cyclic coordinate in the Lagrangian formalism. 
We further choose an appropriate congruence that reaches the critical point. After these steps, the geodesic
equations are analytically solvable in the critical region, and the result can be used to obtain $R$ and $\theta$ as a function of the affine parameter. 

One of the main results of this paper is our assertion that $R \sim \lambda^{-2}$ and $\theta \sim \lambda^{-1}$ in all IG systems with two coupling constants. 
While the validity of this result will be shown in the later sections, we now point out its physical relevance, which can be gleaned as follows. 
We start from the fact that $R \sim \xi^d$ near criticality, as we have already mentioned ($d$ is the spatial dimension of the system). Now, 
$R \sim \lambda^{-2}$ can be translated to $\xi \sim \lambda^{-2/d}$. This is a purely geometric scaling relation for the correlation length, which expresses $\xi$ 
in terms of an affine parameter $\lambda$ close to criticality. $\lambda$, being the square root of the line element, contains information
about all the coupling constants, and is a natural parameter for expressing scaling relations in geometry, as this is a geometric invariant.  

On the other hand, in phase transitions driven by thermal fluctuations, the correlation length exponent is defined by $\xi \sim t^{-\nu}$, where $t = (T - T_c)/T_c$ is 
the reduced temperature. In conjunction with $R \sim \xi^d$,
this implies that $R \sim t^{-\nu d}$ and using the Josephson scaling law \cite{Goldenfeld}, we obtain $R \sim t^{\alpha - 2}$, i.e $R$ has the same scaling dimension as
the correlation volume (see Eq.(6.53) of \cite{rupp}). If the 
coefficient $\alpha = 0$, then $R \sim t^{-2}$. For this case, we will show that $t \sim \lambda$, and our
universal scaling relation is satisfied. However, if $\alpha \neq 0$, then to satisfy $R \sim \lambda^{-2}$, we must have $t \sim \lambda^{\frac{2}{2-\alpha}}$. As we 
show later, this is indeed the case. Importantly, once we obtain the relation between the reduced temperature and the line element, critical exponents
can be obtained in terms of a geometric invariant. 

In the next section, we provide some examples of our method outlined above (further examples demonstrating our results are provided in appendices A to D). 
We will then proceed to give a general mathematical argument to establish the
geometric scaling relations mentioned. We should point out that there are several aspects of IG which are not fully understood from an RG perspective - for example anomalous exponents.
This will include the introduction of a conformal symmetry near criticality. This issue is beyond the scope of the present paper and we will not attempt to discuss it here, 
although this is an important issue for future research. 

\section{A Few Examples}

We illustrate the procedure by the information geometry of the Van der Waals gas for which we set $c_v = 3/2$ to simplify the notation. The Van der Waals model is a characteristic model for
classical liquid-gas like phase transitions, with a first order phase coexistence line culminating in a second order critical point. Here, starting from a standard expression for 
the Helmholtz free energy per unit volume, it is particularly simple to write the metric in the $(T,\rho)$ representation \cite{rupp}, where $\rho = 1/V$. For ease of presentation, 
we set the Boltzmann's constant to unity, and further choose the Van der Waals constants such that the critical temperature and volume, $T_c = V_c = 1$. 
The line element $\mathrm{d}s^2$ is given by (Eqs.(6.58) and (6.59) of \cite{rupp})\footnote{It is important to check the positivity of the line element of IG. From a standard
theorem in mathematics, this implies that (for the two dimensional examples that we consider) the diagonal elements of the metric as well as its determinant is positive. 
In all the examples worked out in this paper, we have checked that this is true in the regions of interest. We do not mention this in sequel.}
\begin{equation}
\mathrm{d}\lambda^2_{VdW}=\frac{3 \rho }{2 T^2}\mathrm{d}T^2
+\frac{9 \left(-\rho ^3+6 \rho ^2-9 \rho +4 T\right)}{4 (\rho -3)^2 \rho  T}\mathrm{d}\rho^2.
\label{vdw}
\end{equation}
The diagonal components of the metric are related to the response functions of the theory. While the first term in Eq.(\ref{vdw}) is proportional 
to $c_v$, the second term is the inverse of the isothermal compressibility. 
This metric leads to the scalar curvature, whose exact expression was calculated in \cite{tapo1}. For our purpose, it is enough to note that this 
diverges as $\left(\rho(\rho - 3)^2 - 4T\right)^{-2}$, which is also the locus of divergence of the isothermal compressibility, and hence defines the spinodal line. 
Close to criticality $T=\rho=1$, we substitute $T = 1 + t$, $\rho = 1 + r$, to obtain $R_{VdW} \sim (3r^2 + 4t)^{-2}$, i.e at criticality, the scalar curvature 
diverges as $t^{-2}$ and is a function of a single variable. Although this is strictly true only along the critical isochore,  it should be true for nearby paths, assuming that
the orders of magnitudes of the fluctuations in $t$ and $\rho$ are the same. Hence, we expect that in this regime, we can write the metric components in terms of one variable. 

To glean further insight into this, we simplify the metric components of Eq.(\ref{vdw}) in the critical regime. 
Retaining the lowest order terms and using a further redefinition of coordinates $x = t - r$, the metric simplifies to
\begin{equation}
\mathrm{d}\lambda^2_{VdW}=\frac{9t}{4}\mathrm{d}x^2-\frac{9t}{2}\mathrm{d}x\mathrm{d}t+\frac{3}{4}\left(2+3t\right)\mathrm{d}t^2.
\end{equation}
Although non-diagonal, this has the advantage that its elements depend on only the coordinate $t$. First, we note that 
the scalar curvature diverges as $R \sim t^{-2}$ as expected. Further, the geodesic equations can 
be obtained from a variational principle from the Lagrangian 
\begin{equation}
{\mathcal L}_{VdW} = \frac{1}{2}\left(\frac{9t}{4}{\dot x}^2 - \frac{9t}{2}{\dot t}{\dot x} + \frac{3}{4}\left(2+3t\right){\dot t}^2\right).
\end{equation}
In particular, since $x$ is a cyclic coordinate, the Euler Lagrange equation corresponding to this gives a first order equation
for ${\dot x}$ in terms of a constant $k_1$. We can solve this simultaneously in conjunction with the normalization condition $u^{\mu}u_{\mu}=1$ to 
obtain the geodesic equations 
\begin{eqnarray}
\dot{x}&=&\frac{4k_1 - \sqrt{6} \sqrt{t \left(9 t-4 k_1^2\right)}}{9t},\nonumber\\
\dot{t}&=-&\frac{1}{3} \sqrt{6-\frac{8 k_1^2}{3 t}}.
\label{geodesics}
\end{eqnarray}
We also set $\lambda = 0$ at criticality, i.e the affine parameter is measured from the critical point. Then, from the above equation, 
we can see that the geodesics won't reach the second order critical 
point $(t,x)=(0,0)$ unless $k_1=0$. Setting this value for $k_1$, we can solve the geodesic equations of Eq.(\ref{geodesics}) analytically, and in particular, we obtain 
$t \sim -\lambda$, where we have also consistently set the constant of integration in the second of Eq.(\ref{geodesics}) to zero (a similar analysis holds for the
geodesic equation involving $x$). This implies that in the critical region,
$R_{VdW} \sim \lambda^{-2}$. The calculation of $\theta$ is straightforward. We construct the tensor $B^{\mu}_{\nu}$ from its definition (recall that 
$B^{\mu}_{\phantom{1}\nu}=\nabla_{\nu}u^{\mu}$) and take its trace. The solution of the geodesic equation is then fed back into this expression in order to obtain $\theta$ as 
a function of $\lambda$. While the first step provides (with $k_1=0$) $\theta_{VdW} \sim  -\frac{1}{t}$, the second gives us $\theta_{VdW} \sim \lambda^{-1}$ close to criticality.
We have thus obtained geometrical critical exponents that describe scaling relations between scalar quantities on the parameter manifold, close to the critical point. 

We have checked that the same qualitative results as above holds in the information geometry of the Curie Weiss model for ferromagnets. This is expected, given that these
systems can be mapped to each other, and shows that our results hold for classical systems with symmetric as well as asymmetric phase diagrams. 

In the models above, the critical exponent corresponding to $c_v$ was zero. Now we comment on a situation in which this ceases to be valid. For simplicity, we 
work in a $(T,\rho)$ representation of IG for single component fluids, for which $g_{TT}$ is proportional to $c_v$ and $g_{\rho\rho}$ is proportional to the 
compressibility. Near criticality, we thus assume a simple form of the metric, in terms of the reduced temperature $t$ : 
\begin{equation}
d\lambda^2 = t^{-\alpha}dt^2 + t^{\gamma}d\rho^2
\end{equation}
where $\alpha$ and $\gamma$ are standard critical exponents \cite{Goldenfeld}.  The scalar curvature computed from this metric is 
$R = -\frac{1}{2} \gamma (\alpha+\gamma-2) t^{\alpha-2}$ (note that this can be simplified to $R = \gamma\beta t^{\alpha - 2}$, using the Rushbrooke scaling 
relation \cite{Goldenfeld}). Since $\rho$ is cyclic, we can follow the procedure outlined in the introduction to obtain 
$u^{\mu} = (-t^{\alpha/2},0)$ for geodesics that reach the critical point. We also calculate the expansion parameter to be $\theta = -\frac{1}{2} \gamma t^{\frac{a}{2}-1}$. 
Now from the form of the vector $u^{\mu}$, we can calculate the dependence of $t$ on the affine parameter $\lambda$, and this can be seen to be 
$t = 2^{\frac{2}{\alpha-2}} ((\alpha-2) \lambda )^{\frac{2}{2-\alpha}}$. Hence, apart from a constant, $t \sim \lambda^{\frac{2}{2-\alpha}}$, as we had expected. 
Now if we substitute this solution in the expressions for $R$ and $\theta$, we obtain 
\begin{equation}
R = -\frac{2\gamma\left(\alpha + \gamma - 2\right)}{\left(\alpha - 2\right)^2\lambda^2},~~~
\theta = \frac{\gamma}{\left(2-\alpha\right)\lambda}
\end{equation}
Thus our claim that $R \sim \lambda^{-2}$ and $\theta \sim \lambda^{-1}$ is established in this case as well, with a non-zero critical exponent for $c_v$.

Next, we apply our method to quantum phase transitions in the transverse XY-spin chain. This is one of the few models that yield analytical results in IG, and
the Hamiltonian with $(2N + 1)$ spins is
\begin{equation}
H_{\rm XY} = -\left[\sum_{j = -N}^{N}\frac{1 + \gamma}{4}\sigma_j^x\sigma_{j+1}^x + \frac{1-\gamma}{4}\sigma_j^y\sigma_{j+1}^y - \frac{h}{2}\sigma_j^z\right]
\label{XYHamil}
\end{equation}
where the $\sigma^i$, $i=x,y,z$ are Pauli matrices, $\gamma$ is an anisotropy parameter and $h$ is an applied magnetic field. 
The Hamiltonian of Eq.(\ref{XYHamil}) can be diagonalized with a series of Jordan-Wigner, Fourier and Bogoliubov transformations. 
The system exhibits two distinct types of phase transitions where the energy spectrum becomes gapless. The transition at $|h|=1$ is an Ising transition from an ordered 
ferromagnetic to a disordered paramagnetic phase, and that for $\gamma = 0, |h| <1$ corresponds to an anisotropy transition between two ordered ferromagnetic phases. 

From an IG perspective, $h$ and $\gamma$ can be thought of as coordinates on the parameter manifold, tuning which can bring the system close to a phase transition, whose
physics is dictated by quantum fluctuations at zero temperature. 
The metric on the PM can be computed exactly following \cite{zan1}, and has a simple form in the region $|h| < 1$, in the thermodynamic limit. 
We will mostly focus on the second order critical line in this region, where 
\begin{equation}
\mathrm{d}\lambda^2_{XY}=\frac{1}{16\gamma(1-h^2)}\mathrm{d}h^2
+\frac{1}{16\gamma(1+\gamma)^2}\mathrm{d}\gamma^2.
\label{xy1}
\end{equation}
We can make the substitution $h={\rm cos}(x)$ and get 
\begin{equation}
\mathrm{d}\lambda^2_{XY}=\frac{1}{16\gamma}\mathrm{d}x^2
+\frac{1}{16\gamma(1+\gamma)^2}\mathrm{d}\gamma^2.
\label{xy2}
\end{equation}
This shows that the metric is regular at $ h=\pm 1 $ (in the absence of any curvature singularity, the geodesics do not show any special behavior there). We can write the geodesic equations
\begin{equation}
\dot{x}=16k_2 \gamma,~~~~\dot{\gamma} = -4 (\gamma +1) \sqrt{\gamma -16 k_2^2 \gamma ^2},
\end{equation}
where $k_2$ is an integration constant that appears in the Euler Lagrange equation for the geodesics, since the metric components of Eq.(\ref{xy2}) are
independent of $x$. $k_2$ will be set to zero for geodesics that reach criticality. Doing this, near criticality, it is seen that $\gamma \sim \lambda^2$. 
It can also be checked that in the critical region, $R_{XY}\sim \gamma^{-1}$ and $\theta_{ XY} \sim \frac{1}{\sqrt{\gamma}}$.
which implies that $R_{XY}\sim \lambda^{-2}$ and $\theta_{XY} \sim \lambda^{-1}$, i.e we obtain the same geometric critical exponents as in CPTs. 

The importance of this result can be understood as follows. Parameter manifolds in classical and quantum phase transitions arise due to entirely different physics.
While the former arises in the context of thermal fluctuations \cite{rupp}, the origin of the latter is quantum in nature \cite{pv}. However, the geometric description of these phase transitions 
do not distinguish between the widely different origins of the Riemannian structures, and hence may be expected to indicate some universal features. Our results indicate that at least for a class of 
systems where the scalar curvature diverges in the critical region, this is indeed true, i.e the scalar curvature and the expansion parameter of a geodesic congruence on the parameter 
manifold show universal scaling behavior near criticality, in terms of a geometric invariant.

\section{A Mathematical Argument}

Admittedly, we have constructed the scalar curvature and the expansion parameter in only a few examples. In appendices A to D, we provide four more examples to 
strengthen our result. These contain similar analyses as presented above on the infinite Ising ferromagnet, the 1-D Ising model, the transverse XY spin chain with an additional
angular parameter, and a somewhat different example arising in gravitational theories. 

However, we now provide a mathematical argument that our results should hold in general whenever there is a curvature singularity in the information geometric 
parameter manifold \cite{curvsing}. To this end, we note that tensor manipulations in Eq.(\ref{evolution}) lead to the celebrated Raychaudhuri equation \cite{poisson}\cite{sayan1}, 
which in two-dimensions (in the absence of shear and rotation parameters) reduce to \cite{reason}
\begin{equation}
{\dot \theta} + \theta^2 + \frac{1}{2}R = 0. 
\label{Raychaudhuri}
\end{equation}
It can be checked that in all the cases that we have considered, the Raychaudhuri equation is satisfied. Now, we assume that the divergence
of $R$ has a power law behavior with $\lambda$ near a curvature singularity, i.e a critical point. This is generally justified, since physical metrics are generically algebraic functions of 
the coordinates near singularities, and in all examples that we have studied, the dependence of $R$ reduces to that on a single coordinate at criticality which can be solved as
a function of $\lambda$ via the geodesic equations. 

Importantly however, this assumption excludes a class of examples, namely those which have 
a delta function singularity in the scalar curvature. These might typically arise from discontinuities in the metric, and result in conical defects (for an example, see \cite{polkovnikov}). 
We are aware of literature in gravitational theories which deal with such singularities, but the nature of these singularities have not been explored in IG. In particular, the
behavior of the Raychaudhuri equation near such a conical singularity in IG is an important topic, and should be studied. We will however not attempt this in this paper. 

Apart from conical defects, i.e in the cases where $R$ has a power law singularity, it is seen from Eq.(\ref{Raychaudhuri}) that $\theta$ should also have a power law behavior near criticality. Also,
$\theta$ is naturally interpreted as the quantity $\frac{1}{l}\frac{\partial l}{\partial\lambda}$, where $l$ is a length that contains all the geodesics in a congruence \cite{poisson}. Taking
$\theta = \beta/\lambda^m$, the definition of $\theta$ implies that for $m < 1$, $l$ goes to a constant as $\lambda \to 0$.
This case is not interesting, as the geodesic congruence does not converge. On the other hand, for $m >1$, we obtain $l \sim {\rm exp}(-\frac{\beta}{\delta\lambda^{\delta}})$ with $\delta = m-1$, 
which would imply an exponential convergence of the congruence near criticality. Although this case cannot be apriori ruled out, we have carried out extensive analysis with different algebraic forms 
of two dimensional metrics, and could not obtain any example where the geodesic congruence converges exponentially fast. We believe that exponential convergence of geodesics 
is possibly unphysical, and it would be potentially interesting to prove this in general. This leaves us with the marginal case $m=1$, for which $\theta \sim \lambda^{-1}$ and we 
see from Eq.(\ref{Raychaudhuri}) that $R \sim \lambda^{-2}$. It therefore seems that the geometric critical exponents are universal for CPTs and QPTs in 2-D PMs, whenever there is a curvature 
singularity. Specifically, if $R \sim A/\lambda^2$ and $\theta \sim B/\lambda$ near criticality with $A$ and $B$ being constants, the Raychaudhuri equation imposes the constraint $A = 2B(1-B)$. 

Importantly, if $B = 1$, then $A = 0$ implies that $R$ can be finite even if $\theta$ diverges near criticality. An example of this situation is provided in Appendix C. 
Let us further elaborate on this point - a divergent behavior of $\theta$ may not necessarily result from a curvature singularity. The simplest example is the two-sphere, where geodesic great
circles converge at the two poles although the sphere has a constant curvature. Thus, a singular behavior of $\theta$ may be simply due to the shape of the PM. To see this in our case, 
it is instructive to consider the transverse XY model with the addition of a rotation parameter $\phi$ which corresponds to rotating all the spins about a $z$-axis \cite{polkovnikov}. 
While the energy spectrum remains the same as the original model with $\phi = 0$, the ground state wave function changes for non-zero $\phi$ and there are additional metric components. 
In the $\gamma - \phi$ plane, for $|h| <1$, the scalar curvature of this model has a delta function singularity at $\gamma = 0$. We find that in this case
the expansion parameter diverges, i.e $\theta \sim \lambda^{-1}$ as $\gamma \to 0$, however, we are unable to comment on this further. However, a similar analysis
in the $\gamma - \phi$ plane for $|h| >1$, for which the PM does not have any singularity reveals that here also $\theta \sim \lambda^{-1}$, as $\gamma \to 0$. In the second case, 
the divergent behavior of $\theta$ is simply due to the spherical shape of the parameter manifold. This is the case $B=1$ of the 
previous paragraph, as we elaborate upon in Appendix C. Thus, whenever a singularity is present 
in the PM indicating a second order phase transition, $\theta$ shows divergent behavior near criticality, for an appropriately chosen congruence. 
The converse is however not true in general. However, note that the Raychaudhuri equation, being a first principles derivation, is always valid. 

\section{Summary and Discussions} 

To summarize, in this paper, we have defined geometric critical exponents in information geometry for systems that exhibit continuous second order classical and quantum 
phase transitions, whenever the scalar curvature of the parameter manifold diverges. These in turn give rise to novel scaling exponents of the correlation length in terms 
of scalar quantities on the parameter manifold, near criticality. As we have pointed out, all critical exponents for temperature driven classical phase transitions
can be written in terms of a geometric invariant, using our results. 

In particular, we have obtained the exponents for the scalar curvature $R$ and the geodesic expansion parameter $\theta$, in terms of
the invariant line element $\lambda$. This was done by appropriately approximating the metric components, which allowed us to solve the 
geodesic equations analytically, close to criticality. We have provided evidence that although the nature of classical and quantum phase transitions are widely different, 
information geometry does not distinguish between their origin and these exponents are possibly universal. Our analysis of geometric exponents indicate 
characteristic properties of the underlying parameter manifolds. We note here that in a recent paper \cite{polkovnikov}, several proposals (see section IV of \cite{polkovnikov}) have been put
forward to experimentally measure the metric on the parameter manifold. If the metric is obtainable even numerically, it might be possible to relate 
our geometric exponents to measurable physical quantities. This might be an interesting direction to explore. 

Although not presented here, we have analyzed geodesics in the geometry of CPTs on the spinodal line, but away from criticality. We find that at these spinodal points, the behavior of $R$ and $\theta$
follow the same exponents as presented here, although this may not be physically relevant, as the system undergoes a first order phase transition before reaching
these spinodal points. We have also studied geodesics in the parameter manifolds in the context of black hole physics (a particular case is presented in Appendix D). 
It is well known that the latter exhibit properties analogous to CPTs in ordinary systems, and our conclusion is that the geometric exponents for these systems again show qualitatively similar
features as indicated in this paper. 

Analysis of geodesics is common in general relativistic systems, where energy conditions
put stringent restrictions on their behavior, via the Raychaudhuri equation \cite{poisson}. However, applications outside the domain of general relativity to generic systems 
like the ones discussed here have not been common, although we are aware of literature that exists on the subject \cite{sayan}. Our results complement these and other
existing results that apply geometric methods to condensed matter systems (see, e.g.\cite{ma1},\cite{ma2}), and 
introduces novel geometric exponents in the physics of classical and quantum second order phase transitions. 

Finally, we point out a few issues that we have not addressed in this paper. Firstly, as we have said, analysis of IG in the context of the renormalization group is
not fully established. This involves introducing a conformal symmetry in the vicinity of the critical point, and we hope to report on this in the near future. Further, 
we have not addressed the issue of delta function divergences in IG, which seem to be relevant in the context of QPTs. This is an important area that needs further investigation. 

\begin{center}
{\bf Acknowledgements}
\end{center}
It is a pleasure to thank Amit Dutta and Sayan Kar for very helpful comments. TS also acknowledges a useful discussion with V. Subrahmanyam. We also thank our
referees for their valuable suggestions to improve various aspects of this work. 

\section{Appendix A : The Infinite Ising Ferromagnet}

The infinite-range ferromagnetic Ising model with a transverse magnetic field has been studied by \cite{diptiman}. The Hamiltonian for this model of $N$ spin $1/2$s is 
\begin{equation}
H_{\rm IIF} = -\frac{J}{N}\sum_{i<j} S_i^zS_j^z - h\sum_iS_i^x = -\frac{J}{2N}\left(S_{\rm tot}^z\right)^2 - hS_{\rm tot}^x
\end{equation}
where $J$ will be set to unity, and the total spin $S_{\rm tot}^z = \sum_iS_i^z$, $S_{\rm tot}^x = \sum_iS_i^x$ (and we have neglected a
constant term). In a mean-field analysis, where the average magnetization $m = \sum_i<S_i^z>/N$, the Hamiltonian for a single spin is 
$H_{\rm IIF}^1 = -mS_{\rm tot}^z - hS_{\rm tot}^x$. This is a two-state model whose partition function can be shown to be given by \cite{diptiman}
\begin{equation}
Z = 2{\rm Cosh}\left(\frac{\sqrt{h^2 + m^2}}{2T}\right)
\end{equation}
Geometric aspects of this model have been discussed in \cite{tapo3a} and we simply state the metric elements 
\begin{eqnarray}
g_{TT} &=& \frac{1}{4T^4}\left(h^2 + m^2\right){\rm Sech}^2\alpha \nonumber\\
g_{mm} &=& \frac{1}{T} - \frac{1}{4T^2}{\rm Sech}^2\alpha
\frac{\left(m^2\sqrt{h^2 + m^2} + h^2T{\rm Sinh}(2\alpha)\right)}{\left(h^2 + m^2\right)^{3/2}}
\label{iffmetric}
\end{eqnarray}
where $\alpha = \sqrt{h^2 + m^2}/2T$. We focus on a critical point near $m=0$, and work unto linear order in $m$. In this limit, from the form of the metric in Eq.(\ref{iffmetric}), 
the magnetization can be set to zero, and the the scalar curvature 
can be shown \cite{tapo3a} to diverge at ${\rm Tanh}\frac{h}{2T} = 2h$, defining the phase boundary. This matches with the result of \cite{diptiman}. We now choose 
a value $h = 0.2$, for which the critical temperature is $T_c = 0.236$. To understand the nature of geodesics near the critical point $(T=0.236,m=0)$, we start from the 
vector $u^{\mu} = \left(T'(\lambda),m'(\lambda)\right)$. The fact that $m$ is a cyclic coordinate allows us to analytically solve for $T'(\lambda) $ and $m'(\lambda)$. The solution
is obtained in terms of a constant (arising because $m$ is cyclic) which has to be set to zero for geodesics to reach the critical point. Doing this, we obtain
\begin{equation}
u^{\mu} = \left(\frac{2 T^2}{g} \cosh \left(\frac{g}{2 T}\right),0\right)
\end{equation}
The dependence of the temperature on the affine parameter is obtained as 
\begin{equation}
T = 0.05\left(\tanh ^{-1}\left[\tan (0.206\, -0.5 \lambda )\right]\right)^{-1}
\label{tsoliif}
\end{equation}
where in an intermediate step, an appropriate constant has been chosen in the solution so that at $\lambda = 0$, $T = T_c = 0.236$. 
The expansion scalar is obtained by a standard analysis and is given by
\begin{equation}
\theta = \frac{\text{sech}\left(\frac{h}{2 T}\right) \left(T \sinh \left(\frac{h}{T}\right)-4 h T \cosh ^2\left(\frac{h}{2T}\right)+h\right)}{2 h \left(2 h-\tanh \left(\frac{h}{2 T}\right)\right)}
\label{thetaiif}
\end{equation}
Upon substituting the solution of $T$ of Eq.(\ref{tsoliif}) into Eq.(\ref{thetaiif}) and expanding in powers of $\lambda$, we obtain (with $h=0.2$) 
\begin{equation}
\theta = -1.179 +\frac{1}{2\lambda }  -2.945 \lambda + O(\lambda^2)
\end{equation}
The expression for the scalar curvature appears in Eq.(19) of \cite{tapo3a}. If we substitute Eq.(\ref{tsoliif}) into that equation and expand in powers of $\lambda$, we get
\begin{equation}
R = 9.001\, +\frac{1}{2\lambda ^2}  +\frac{2.357}{\lambda } +29.478 \lambda + O(\lambda^2)
\end{equation}
This proves the assertions for the divergences of $\theta$ and $R$ in the main text. It can also be shown that the Raychaudhuri equation is satisfied in this case. 

\section{Appendix B : The 1D Ising Model}

Geometric aspects of the 1D Ising model has been very well studied in the literature. We skip all the details here, for which we refer the interested reader to
\cite{janyszek}, where the metric components and scalar curvature were analyzed. This model is somewhat complicated to handle, and for our analytic methods
to be effective, we will work in the zero field limit. Specifically, with the Hamiltonian 
\begin{equation}
H_{\rm Ising} = -J\sum_{i=1}^N S_iS_{i+1} - h\sum_i^NS_i,
\end{equation}
and defining the variables $x = J/T$ and $y = h/T$ (where we have set the Boltzmann's constant to unity) we will work in the limit 
$e^{4x}{\rm sinh}^2y \ll 1$. In that case, from \cite{janyszek}, it can be shown that the metric elements for the 1D Ising model is given by
\footnote{There is a small subtlety here. The scalar curvature calculated from the metric of Eq.(\ref{metric1dising}) is $R=-(1+e^{2x})$. Its equals the zero field limit
of the expression given in \cite{janyszek}. However, there is a change in sign. Since we are only concerned about the magnitude of $R$, this will not affect our
discussion here.} 
\begin{equation}
g_{xx} = \frac{4e^{2x}}{\left(1 + 2x\right)^2},~~~g_{yy} = e^{2x}.
\label{metric1dising}
\end{equation}
Now denoting the affine parameter as $\lambda$, and following the same procedure outlined before, we obtain  
\begin{equation}
x(\lambda) = \log \left(\cot \left(\frac{\lambda }{2}\right)\right),
\end{equation}
where we have appropriately chosen a constant in the solution so that $x$ diverges at $\lambda = 0$, corresponding to the critical point $T=0$. Noting that
the expansion parameter is given in this case by 
\begin{equation}
\theta = -\frac{1}{2} e^{-x} \left(e^{2 x}+1\right)
\end{equation}
we see that near criticality, $\theta = - {\rm cosec}\lambda$ and $R = -{\rm cosec}\left(\frac{\lambda^2}{2}\right)$. These have the expansions
\begin{equation}
\theta = -\frac{1}{\lambda }-\frac{\lambda }{6} + O(\lambda^3),~~~R = -\frac{1}{3} -\frac{4}{\lambda ^2} + O(\lambda^2)
\end{equation}
We have numerically verified that the Raychaudhuri equation is satisfied in this example. 

\section{Appendix C : XY Spin Chain on $\phi-\gamma$ plane}

In this appendix, we study the XY spin chain in the presence of an additional angular parameter (see section C of \cite{polkovnikov}). 
We focus on the $\phi-\gamma$ plane in the XY spin chain for $h=2$, i.e the case when there are no curvature singularities. The metric is given by
(Eq.(31) of \cite{polkovnikov}) :
\begin{eqnarray}
g_{\gamma\gamma} &=& 
\frac{1}{16} \left[\frac{2}{\left(1-\gamma ^2\right)^2}\left(\frac{h}{\sqrt{\gamma^2+h^2-1}}-1\right)-\frac{\gamma ^2 h}{\left(1-\gamma ^2\right) \left(\gamma^2+h^2-1\right)^{3/2}}\right]
\nonumber\\
g_{\phi\phi} &=& \frac{\gamma ^2}{8 \left(1-\gamma ^2\right)} \left(\frac{h}{\sqrt{\gamma ^2+h^2-1}}-1\right)
\end{eqnarray}
We will choose $h=2$, so that we are in the paramagnetic disordered phase, where there is no singularity in the parameter manifold. In that case, $\phi$ is a cyclic coordinate and
the method developed in the main text can be applied.  First, the scalar curvature is given by 
\begin{equation}
R = -\frac{96 \left[\left(5404 \sqrt{3}-9360\right) \gamma ^2+2967\sqrt{3}-5139\right]}{\left(\left(22 \sqrt{3}-38\right) \gamma ^2+3\left(4 \sqrt{3}-7\right)\right)^3},
\end{equation}
which is regular at $\gamma = 0$. The expansion parameter can likewise be computed to be 
\begin{equation}
\theta = \frac{2 \sqrt{2} \left(2 \sqrt{3}-3\right) \sqrt{2 \left(5\sqrt{3}-9\right) \gamma ^2+6 \sqrt{3}-9}}{\gamma  \left[\left(22\sqrt{3}-38\right) \gamma ^2+3 \left(4 \sqrt{3}-7\right)\right]}
\end{equation}
Now we can solve for $\gamma$ as a function of the affine parameter $\lambda$. The solution is somewhat complicated, and we do not reproduce it here. We mention the
result that in terms of this solution, the the expansion parameter near $\gamma = 0$ can be expressed as
\begin{equation}
\theta = \frac{1}{\lambda} + 12.62\lambda^2 + O(\lambda^3)
\end{equation}
That the Raychaudhuri equation is satisfied can be verified numerically. This is an example of a singularity in $\theta$ without any singularity in $R$ and corresponds
to the coefficient $B=1$ of section 3. 

\section{Appendix D : An example from gravity}

In this final appendix, we consider an example of our construction in a theory of gravity. We consider a Reissner-Nordstrom anti-de-Sitter (RN-AdS) black hole 
in $3+1$ space-time dimensions. These are electrically charged black holes in Einstein-Maxwell theory with charge $Q$. 
These black holes, which possess entropy (denoted by $S$) and temperature, obey laws of black hole thermodynamics have been well
studied in the literature (for a recent analysis, see \cite{tapo5}). Starting from the formula for the black hole mass, given by
\begin{equation}
M = \frac{\sqrt{S} }{2 \sqrt{\pi }}\left(1+\frac{S}{\pi}\right)+\frac{\sqrt{\pi } Q^2}{2 \sqrt{S}}
\end{equation}
where we have set an AdS length scale to unity, one obtains after a fairly straightforward analysis, the metric on the two dimensional space of
parameters $(Q, S)$ reads
\begin{equation}
\mathrm{d}\lambda_{BH}^2=\frac{4 \pi ^2 S}{g(S,Q)}\mathrm{d}Q^2
-\frac{4 \pi ^2 Q}{g(S,Q)}\mathrm{d}S\mathrm{d}Q\\
+\frac{3 \pi ^2 Q^2+3 S^2-\pi  S}{2 S g(S,Q)}\mathrm{d}S^2
\label{RNAdSmetric}
\end{equation}
where $g(S,Q)=\left( 3 S^2 + \pi S - \pi^2Q^2\right)$. It is to be noted that the fluctuating variables here are the charge $Q$ and the mass $M$ when the
black hole is treated in a grand canonical ensemble where it can exchange charge and mass with the surroundings, and $S$ is to be understood as being
a function of $M$ and $Q$, from Eq.(\ref{RNAdSmetric}). This system can be shown to resemble a liquid-gas system, with the spinodal line defined
by the equation $\pi ^2 Q^2+3 S^2-\pi  S=0$, which is also the locus of divergence of the curvature scalar, as follows from calculating $R$, using
Eq.(\ref{RNAdSmetric}). Also the first law of black hole thermodynamics, $dM = TdS + \phi dQ$ fixes the electric potential to be $\phi = Q\sqrt{\pi/S}$. We are
thus interested in identifying the behaviour of $R$ and $\theta$ near the second order critical point. 

The first step in our analysis is to define a new variable $y=\phi^2$, and work with the variables $(S,y)$, in terms of which the off-diagonal terms in the metric vanish, and we have 
\begin{equation}
\mathrm{d}\lambda_{BH}^2=\frac{3 S+\pi  (y-1)}{2 S (3 S-\pi  y+\pi )}\mathrm{d}S^2+
\frac{\pi  S}{y (3 S-\pi  y+\pi )}\mathrm{d}y^2
\end{equation}
The spinodal curve in the new variables satisfies the equation $y=1-3S/\pi$. Now, the location of the critical point can be seen, from the maximum of the spinodal
curve, as $S_c = \pi/6$, $y_c = 1/2$, corresponding to $Q_c = 1/(2\sqrt{3})$. We expand this metric upto first order, by setting $S = S_c + s, y = y_c + h$. The resulting
metric is simple, but still difficult to solve analytically. We thus make a further change of variables, and define the new coordinate $x = \pi h + 3s$. Upon following these steps,
the metric near criticality finally reduces to 
\begin{equation}
\mathrm{d}\lambda_{BH}^2=\frac{x}{3\pi^2}\mathrm{d}x^2-\frac{2x}{3\pi}\mathrm{d}x\mathrm{d}h
+\frac{\pi +x}{3}\mathrm{d}h^2
\label{RN3}
\end{equation}
Upon using the cyclic coordinate $h$ and the normalization condition, we obtain the solution for $u^{\mu}$ in terms of a constant $k$ as
\begin{equation}
\dot{x}=-\sqrt{3 \pi } \sqrt{\frac{-3 k^2+x+\pi }{x}}, ~~~\dot{h}=\frac{3 \pi  c-\sqrt{3 \pi } \sqrt{x \left(-3 k^2+x+\pi \right)}}{\pi  (x+\pi )}. 
\end{equation}
For geodesics which reach infinitesimally close to criticality, we will need to set the constant $k = \sqrt{\pi/3}$. Using this fact, we compute the expansion 
parameter to be $\theta = -\sqrt{3\pi}/(2x)$. Further, we calculate the dependence of $x$ on the affine parameter $\lambda$, and find that this is given by
$x = -\sqrt{3\pi}\lambda$. This implies that $\theta = 1/(2\lambda)$. The curvature scalar that follows from Eq.(\ref{RN3}) is $R = 3\pi/(2x^2)$, which now gives
$R = 1/(2\lambda^2)$. These expressions can be seen to be consistent with the results in the main text. The Raychaudhuri equation can also be seen to
be satisfied from these relations.


\begin{thebibliography}{99}
\bibitem{curvature}
$R = \frac{2}{\sqrt{g}}\left[\frac{\partial}{\partial x^2}\left(\frac{\sqrt{g}}{g_{11}}\Gamma^2_{11}\right) -\frac{\partial}{\partial x^1}\left(\frac{\sqrt{g}}{g_{11}}\Gamma^2_{12}\right)\right]$,
where $g$ is the determinant of the metric, and the Christoffel symbols are defined by 
$\Gamma^{\mu}_{\nu\rho} = \frac{1}{2}g^{\mu\zeta}\left(\frac{\partial g_{\zeta\nu}}{\partial x^{\rho}} + \frac{\partial g_{\zeta\rho}}{\partial x^{\nu}}
- \frac{\partial g_{\nu\rho}}{\partial x^{\zeta}}\right)$.
\bibitem{geofocus}
Note however that the converse is not true : geodesics focusing at a point does not necessarily indicate a singularity there. This is most easily illustrated by considering a two-sphere, 
where great circles focus at the poles where there are no singularities.
\bibitem{curvsing} When this is not the case, for example along the $h=1$ transition line in the XY-spin chain model where the scalar curvature does not show a divergence, 
our method will not be applicable. For these cases, geodesics do not show any special behavior near the critical region.
\bibitem{reason} This derivation is based on the fact that in two dimensions, the elements of the Ricci tensor are related to the metric by 
$R_{\mu\nu} = \frac{1}{2}g_{\mu\nu}R$.
\end{thebibliography}
\end{document}